\documentclass[twocolumn,letter]{jpsj3} 
\title{Evolution toward Quantum Critical End Point in UGe$_2$}

\author{Hisashi \textsc{Kotegawa}$^{1,2}$\thanks{E-mail address: kotegawa@crystal.kobe-u.ac.jp}, Valentin \textsc{Taufour}$^2$, Dai \textsc{Aoki}$^2$, Georg \textsc{Knebel}$^2$, Jacques \textsc{Flouquet}$^2$}

 \inst{$^1$Department of Physics, Kobe University, Kobe 657-8501, Japan\\
$^2$INAC/SPSMS, CEA-Grenoble, 17 rue des Martyrs, 38054 Grenoble, France

}

\abst{
We report on Hall resistivity and electrical resistivity measurements under pressure and magnetic field in UGe$_2$ with ferromagnetic (FM) tricriticality.
The Hall resistivity sensitively detects the first order metamagnetic transition from a paramagnetic (PM) phase to a FM phase in a large pressure range almost up to the quantum critical end point (QCEP).
The drastic change in the Fermi surface at the PM--FM transition is detected up to the vicinity of the QCEP, while a strong modification in the field variation of the inelastic scattering between electrons is observed toward the QCEP.
The comparison with the theoretical predictions is made.

}

\kword{quantum critical end point, UGe$_2$, metamagnetism, Hall effect}

\begin{document}
\maketitle

The Curie temperature $T_{\rm Curie}$ of a paramagnetic (PM) - ferromagnetic (FM) phase transition can be driven to 0 K by applying a critical pressure $P_{\rm C}$.
In itinerant FM systems, the PM--FM transition goes from second order to first order at the tricritical point (TCP) which is at a pressure below $P_{\rm C}$.\cite{Pfleiderer_MnSi,Uhlarz,Kabeya,Valentin,Belitz,Yamada}
One consequence of the change of order at the TCP is that, above $P_{\rm C}$ and at low temperature, the first order PM--FM transition is induced by the magnetic field along the easy magnetization axis.
At higher temperature, this first order transition changes to a crossover through a critical end point at $T_{\rm CEP}$.
The line of $T_{\rm CEP}$ under pressure and magnetic field defines FM wings.\cite{Valentin,Belitz}
Such behavior would not be observed for a second order phase transition since the applied magnetic field in itself breaks the time reversal symmetry.
On increasing pressure, the field induced first order PM--FM transition terminates at a quantum critical end point (QCEP) at $0\,{\rm K}$ which is characterized by its pressure $P_{\rm QCEP}$ and its field $H_{\rm QCEP}$.
Thus $T_{\rm CEP}$ starts at $T_{\rm TCP}$ at $H=0$ and ends up at $H_{\rm QCEP}$ at $T=0$ K.
The QCEP is clearly distinguished, by the lack of spontaneous symmetry breaking, from a conventional quantum critical point which is a second order phase transition at 0\,K.
The QCEP is a fascinating target for theories \cite{Belitz,Yamada,Millis,Binz,Yamaji,Imada} but experimentally represents a challenge due to the lack of systems which can be studied at accessible field and pressures.
Here as the FM wing is drawn almost up to QCEP, comparison can be made with theoretical proposals.

UGe$_2$ is a unique example as the conventional itinerant FM limit ($P_{\rm C} \gg 0$) has been precisely determined and it is possible to produce high quality single crystals.
Furthermore, UGe$_2$ has been already extensively studied as it is the first discovered ferromagnetic superconductor.\cite{Saxena}
At ambient pressure, the Curie temperature ($T_{\rm Curie}\sim 52\,{\rm K}$) is associated with the large sublattice magnetization $M_0 \sim 1.5\,\mu_{\rm B}$.
With increasing pressure, the magnetic ground state switches from a highly polarized phase (FM2, $M_0 \sim 1.5\,\mu_{\rm B}$) to a weakly polarized phase (FM1, $M_0\sim 0.9\,\mu_{\rm B}$) at a pressure $P_x\sim1.2\,{\rm GPa}$ through a first order transition.
Further increasing the pressure, the FM1 phase collapses at a critical pressure $P_{\rm C}\sim 1.5\,{\rm GPa}$ with an abrupt drop of the sublattice magnetization $\Delta M_0=0.9\,\mu_{\rm B}$.\cite{Pfleiderer}
De facto, it is this large drop of $\Delta M_0$ which allows one to observe the pressure and magnetic field evolution of the transitions over a large pressure-field window up to QCEP.
Recently the TCP at zero field has been observed by thermal expansion measurements \cite{Kabeya} and resistivity measurements which indicate $P_{\rm TCP}=1.42\,{\rm GPa}$ and $T_{\rm TCP}=24\,{\rm K}$.\cite{Valentin}

Here we report the pressure and field evolution of PM--FM1 transition by
Hall and resistivity measurements up to $3.41\,{\rm GPa}$ in UGe$_2$.
In our previous experiments,\cite{Valentin} we reported the observation of the TCP and the ($P,T,H$) phase diagram with the first order plane.
However, the pressure was much lower than $P_{\rm QCEP}$.
Other previous measurements were also limited to 2 GPa.\cite{Haga,Kobayashi,Pfleiderer,Terashima} 
In the present experiments we were able to approach close to the QCEP and demonstrate the clear separation between the first order transition and crossover regime by Hall measurements which have the benefit of being sensitive to changes in Fermi surface (FS).

A single crystal of $0.8 \times 0.3 \times 0.15\, {\rm mm}^3$ was used and was prepared as previously described.\cite{Valentin}
The Hall resistivity $\rho_{xy}$ and electrical resistivity $\rho_{xx}$ were measured using a four-probe AC method.
The electrical current was applied along the $c$-axis in the orthorhombic structure.
 $\rho_{xy}$ was extracted from the difference between positive and negative field, namely $(\rho_{H+}-\rho_{H-})/2$, while the $\rho_{xx}$ was obtained by the average, namely $(\rho_{H+}+\rho_{H-})/2$.
Pressure was applied up to $3.41\,{\rm GPa}$ using an indenter cell with Daphne7474 as a pressure-transmitting medium.\cite{Indenter,Murata}
The pressure was determined by the superconducting transition temperature of lead.
For low pressures, up to $2.72\,{\rm GPa}$, the measurements were performed using Quantum Design Physical Properties Measurement System (PPMS) with a maximum field of 9 T and a base temperature of $1.8\,{\rm K}$,
while for higher pressures a conventional dilution refrigerator was used with a maximum field of $16\,{\rm T}$ and a base temperature of $0.22\,{\rm K}$.

\begin{figure}[htb]
\centering
\includegraphics[width=0.7\linewidth]{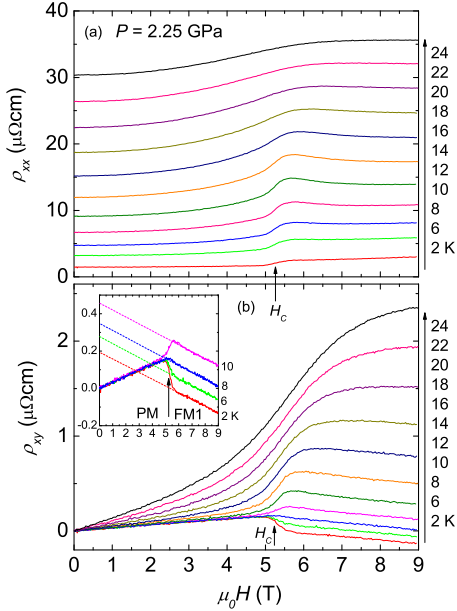}
\caption[]{(color online) Field dependence of (a) $\rho_{xx}$ and (b) $\rho_{xy}$ for $H \parallel a$-axis at 2.25 GPa. The transitions at $H_{\rm C}$ are of first order at low temperatures, and change to a crossover at high temperatures. The inset in lower panel shows $\rho_{xy}$ at low temperatures. The finite intercept extrapolated from the FM1 phase above $H_{\rm C}$ confirms that FM1 phase has a spontaneous magnetization.
}
\end{figure}

Figure 1 shows the field dependence of $\rho_{xx}$ and $\rho_{xy}$ for several temperatures at 2.25 GPa, where the ground state of UGe$_2$ is PM at zero field.
When the magnetic field is applied along the $a$ axis (easy magnetization axis) at low temperature, the system shows the metamagnetic transition from the PM phase to the FM1 phase at $H_{\rm C}$ accompanied by a step-like jump in $\rho_{xx}$ and $\rho_{xy}$.
At higher temperatures, the PM--FM1 transition changes to a crossover.\cite{Valentin}
In both, $\rho_{xx}$ and $\rho_{xy}$, the sharp anomaly appears at $H_{\rm C} \sim 5.2$ T at low temperatures, but it is smeared out at high temperatures.
The inset of Fig.~1(b) shows $\rho_{xy}$ between $2$ and $10$\,K.
The field dependence of $\rho_{xy}$ is almost linear for both the PM and the FM1 phase.
In general, $\rho_{xy}$ is expressed as follows,
\begin{equation}
\rho_{xy} = R_0 \mu_0 H + R_s \mu_0 M,
\end{equation}
where $M$ is the magnetization.
The first term is attributed to the ordinary Hall effect related to the carrier density and the carrier mobility.
The second term is the anomalous Hall effect originating from skew scattering and side-jump scattering, and it has been reported that the relation of $R_S \propto \rho_{xx}^2$ is dominant in the FM state of UGe$_2$.\cite{Tran}
At low temperatures, no strong field dependence of $\rho_{xx}$ is observed in either the PM or the FM1 phase hence the linear field dependence of $\rho_{xy}$ observed in these phases which indicates that field dependence of the carrier density is small.
The sign change in the slope between FM1 and PM implies a change in the FS with opposite signs for the dominant carrier.
At ambient pressure, band structure calculations have shown that there is a topological change at the FS of UGe$_2$, which is a compensated metal, between 2 hole bands and 2 electron bands in the FM phase and 2 hole bands and 1 electron band in the PM phase.\cite{Settai}
The intercept of $\rho_{xy}$ extrapolated to $H=0$ corresponds to $R_s \mu_0 M(H\rightarrow 0)$.
This term is zero in the PM state while it has a nonzero value in the extrapolation from the FM1 phase, indicative of a spontaneous magnetization.
The quantitative analysis is difficult due to the lack of absolute magnetization data under pressure, but the Hall effect is found to be a good tool to detect the PM--FM transition through the changes in both the FS and in $M$.

\begin{figure}[htb]
\centering
\includegraphics[width=0.7\linewidth]{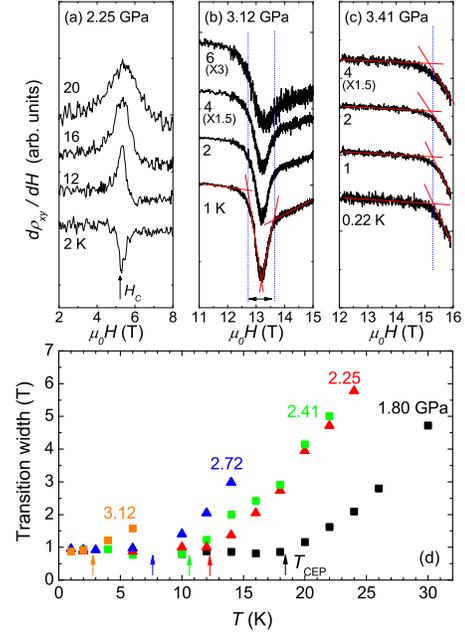}
\caption[]{(color online) Field dependence of $d \rho_{xy} / d H$ at (a) 2.25, (b) 3.12, and (c) 3.41 GPa. $H_{\rm C}$ was deduced from the peak in $d \rho_{xy} / d H$. The transition width was estimated from the cross point of the linear extrapolations. (d): Temperature dependence of the transition width in the field sweep. The width is kept as sharp as $\sim1$ T in the first-order transition, but it is broadened in the crossover regime at high temperatures.  The $T_{\rm CEP}$ between first-order and crossover is indicated by arrows.
}
\end{figure}

Figure 2 shows the field dependence of $d\rho_{xy}/dH$ at 2.25, 3.12, and 3.41 GPa.
Clear maxima or minima were observed due to the PM--FM1 transition which is defined as $H_{\rm c}$.
As shown in Fig. 2(b), the transition width, indicated by the arrows, was estimated using the cross point of linear extrapolations. At 2.25 GPa, the transition width is unchanged below $12\,{\rm K}$,
although the anomaly changes from a maximum to a minimum.
Above $12\,{\rm K}$ the anomalies become broader with the increase of transition width.
Figure 2(d) shows the temperature variation of the transition width for various pressures.
The transition width at low temperature is approximately $1\,{\rm T}$ and almost temperature independent, but it broadens when entering the crossover regime by increasing the temperature.
The boundary between the first order transition and the crossover is indicated by arrows and  allows us to determine $T_{\rm CEP}$.
The present results are in good agreement with the previous results obtained by $d\rho/dT$.\cite{Valentin}
For example, the previous estimation by $d\rho/dT$ indicates $T_{\rm CEP}$ of 18 K at 1.82 GPa,
while the present results obtained by $d\rho_{xy}/dH$ indicates $T_{\rm CEP}$ of 18 K at 1.80 GPa.
As shown in Fig.2, $H_{\rm C}$ increases and $T_{\rm CEP}$ decreases with increasing pressure.
At 3.12 GPa, $T_{\rm CEP}$ is approximately 3 K, and $H_{\rm C}$ is 13.2 T.
At 3.41 GPa, $H_{\rm C}$ exceeds our maximum field of 16 T and only the beginning of the PM--FM1 transition is detected.
We roughly estimated $T_{\rm CEP} = 1.5 \pm 1 \,{\rm K}$ and $H_{\rm C} = 16.5 \pm 0.5\,{\rm T}$ at 3.41 GPa from comparison with data at other pressures.

\begin{figure}[tb]
\centering
\includegraphics[width=\linewidth]{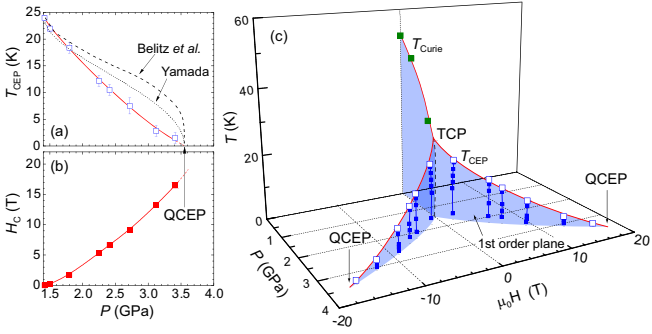}
\caption[]{(color online) Pressure dependence of (a) $T_{\rm CEP}$ and (b) $H_{\rm C}$ at low temperature. The QCEP is expected to be located at $3.5-3.6$ GPa and $17-19$ T from these plots. The dashed and dotted curves are the theoretical predictions by Belitz {\it et al.} and Yamada obtained using ($T_{\rm TCP}=24$~K, $P_{\rm QCEP}=3.55$~GPa, $H_{\rm QCEP}=18$~T).\cite{Belitz,Yamada}  (c): Three dimensional phase diagram of UGe$_2$.

}
\end{figure}

Figures 3 show the pressure dependence of  $T_{\rm CEP}$ and  $H_{\rm C}$ at the lowest temperature.
$T_{\rm CEP}$ decreases with pressure and the variation becomes weaker at higher pressure,
while the slope of $H_{\rm C}$ becomes steeper.
Figure 3(c) shows the three dimensional phase diagram drawn from the present results.
$T_{\rm CEP}$ becomes zero at $P_{\rm QCEP} \sim 3.5-3.6$ GPa and $H_{\rm QCEP} \sim 17-19$ T from the extrapolations shown in Figs.~3(a) and (b).

\begin{figure}[htb]
\centering
\includegraphics[width=\linewidth]{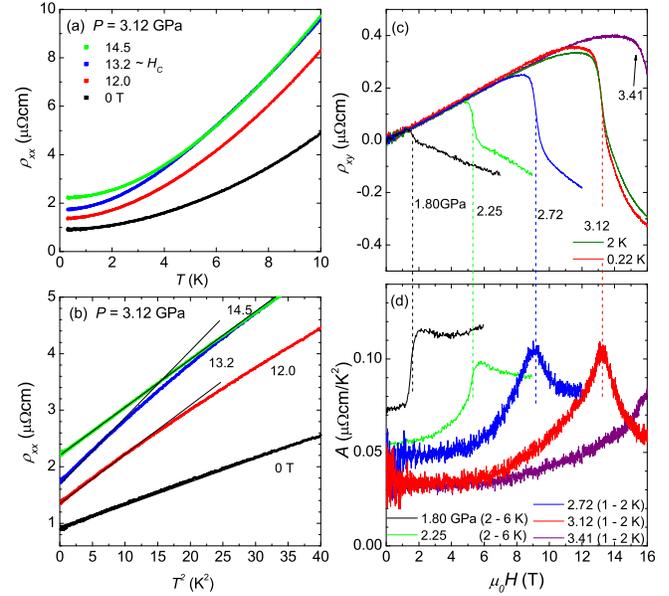}
\caption[]{(color online) (a) Temperature dependence of $\rho_{xx}$ and (b) $\rho_{xx}$ vs. $T^2$ at 3.12 GPa, where $H_{\rm C}$ is 13.2 T and $T_{\rm CEP}$ is estimated to be $\sim3$ K from Fig.~2(d). Field dependence of (c) $\rho_{xy}$ at low temperatures, and (d) $A$ coefficient at several pressures. $\rho_{xy}$ for $1.80 - 2.72$ GPa was measured at 2 K, and 3.41 GPa is 0.22 K. The values of the $A$ coefficient were obtained from the field dependence of $\rho_{xx}$ at two denoted temperatures by assuming the FL form. The dotted lines indicate $H_{\rm C}$ at each pressure.
}
\end{figure}

Figure 4(a) shows temperature dependence of $\rho_{xx}$ for several fields at 3.12 GPa close to $H_{\rm QCEP}$ where $T_{\rm CEP}$ and $H_{\rm C}$ are $\sim3$ K and 13.2 T respectively.
No clear anomaly was observed in $\rho_{xx}$ at $T_{\rm CEP}$ within the experimental precision, although the anomaly is observed when $T_{\rm CEP}$ is high at lower pressures.\cite{Valentin}
Figure 4(b) shows the $T^2$ plot of $\rho_{xx}$.
At 0 T and 14.5 T, the Fermi liquid (FL) behavior of $\rho_{xx} = \rho_0 +AT^2$ is observed in a wide temperature range, while $\rho_{xx}$ obeys the FL behavior below approximately $T_{\rm CEP} \sim 3$ K at 12 T and 13.2 T.
The $A$ coefficient in the FL region is the largest at $H_{\rm C} \sim 13.2$ T, indicative of an enhancement of quasiparticle mass on crossing $H_{\rm C}$.
The FL regime shrinks on approaching $H_{\rm C}$, but careful analysis is required to discuss the origin of this behavior, since the residual resistivity $\rho_0$, which is different between the PM phase and the FM1 phase, is expected to depend on temperature especially near $H_{\rm C}$ in a wide temperature range from the crossover region at high temperatures to the first order region at low temperatures.

Figure 4(c) shows the field dependences of $\rho_{xy}$ at low temperatures and (d) of the $A$ coefficient estimated from the field dependences of $\rho_{xx}$ at two different temperatures by assuming the FL form of $\rho_{xx} = \rho_0 +AT^2$.
$\rho_{xy}$ shows a step-like anomaly at $H_{\rm C}$ up to 3.12 GPa, indicating that the drastic change in the FS is maintained up to the vicinity of the QCEP.
On the other hand, the field dependence of $A$ shows a large evolution toward the QCEP.
At lower pressure with high $T_{\rm CEP}$, $A$ shows a step-like anomaly at $H_{\rm C}$ which is 
indicative of a first-order transition between two phases with different FS's and effective masses.
This behavior is consistent with the previous report.\cite{Terashima}
The step-like anomaly of $A$ gradually changes into peak structures with increasing pressure.
Both $A$ at zero field and $A$ at FM1 sufficiently above $H_{\rm C}$ decrease with increasing pressure, while $A$ at $H_{\rm C}$ is almost pressure independent.
It is apparent that the ratio $A(H_{\rm C})/A(H=0)$ increases toward the QCEP.
This strong modification in the mass enhancement may be driven by the continuous decrease of $\Delta M_0$ on approaching QCEP at the benefit of a large increase in magnetic fluctuations.
Similar enhancement of $A$ under magnetic field has been observed in antiferromagnetic (AF) heavy fermion compounds such as CeRh$_2$Si$_2$ on crossing its first order metamagnetic transition\cite{Knafo} or CeIn$_3$ and YbRh$_2$Si$_2$ on crossing its magnetic transition from AF to PM.\cite{Silhanek,Gegenwart}
For the two cases, the magnetic transition is associated with FS changes.
For example, in CeIn$_3$ it was proposed that it is associated to a Lifshitz instability,\cite{Gor'kov} and in YbRh$_2$Si$_2$ the anomaly in Hall coefficient was interpreted as a collapse of the large FS.\cite{Paschen}

Studies on other FM materials such as ZrZn$_2$ suffer from the fact as the jump $\Delta M_0$ is one order of magnitude smaller than that of UGe$_2$~\cite{Uhlarz} the separation between $P_{\rm QCEP}$ and $P_{\rm C}$ seems very small and thus the FM wing has been only drawn schematically.
The enhancement of the $A$ coefficient at $H_{\rm C}$ is also observed in Sr$_3$Ru$_2$O$_7$~\cite{Grigera} which is considered to lie in the vicinity of FM QCEP.
Unfortunately Sr$_3$Ru$_2$O$_7$ is already in PM ground state and thus the validity of a dominant FM coupling is not obvious.
It was stressed that an AF instability may play an important role~\cite{Kitagawa} with a possible duality between FM and AF tricriticality~\cite{Misawa} in a manner similar to the highly studied CeRu$_2$Si$_2$ family~\cite{Flouquet_review_2006,Aoki}.
Finally, MnSi suffers from the complexity that due to its lack of an inversion symmetry, the ordered ground state at zero field is not a FM.\cite{Pfleiderer_MnSi}
UGe$_2$ is an excellent system with remarkable separation between $P_{\rm QCEP}$ and $P_{\rm C}$.

Because of the large pressure and field extensions of the first-order plane, UGe$_2$ is a good test for theories of the FM phase diagram.
Belitz {\it et al.} argued that long wavelength correlation effects can explain
such a phase diagram with the wing structure.\cite{Belitz}
In Belitz's model which gives $T_{\rm CEP}$ vs. $H_{\rm C}$, we obtained the pressure dependence of $T_{\rm CEP}$ using the pressure dependence of $H_{\rm C}$.
However, as shown by the dashed line in Fig.~3(a), the quantitative agreement with our data is poor.
Yamada explained the pressure dependence of $T_{\rm CEP}$ by taking into account the magnetoelastic coupling.
In Yamada's model, we used $q(0)=0.16$ and $\eta=0.01$ as parameters in ref.\cite{Yamada}, where $\eta$ connects to the magnetovolume coupling and the result has no large $\eta$ dependence.
However again the predicted $T_{\rm CEP}$ variation of $(P-P_{\rm QCEP})^{1/2}$ does not seem to agree with our results.
These discrepancies should inspire more refinements in theories.
For example, it would be important to take into account the change of FS at the PM--FM1 transition which has been clearly detected by quantum oscillation experiments \cite{Settai} and the present Hall resistivity measurements.
The FS's of the phases FM2, FM1 and PM are quite different.\cite{Haga,Settai}
Thus an interesting problem is the role of FS change on the quantum metamagnetic transition.
As pointed out in the hypothesis of a so-called Lifshitz transition, unconventional universality may occur at quantum metamagnetic transition.\cite{Yamaji,Imada}
The discrepancy between our results and the conventional description of the FM wing structure appears as the signature of a FS change at the PM--FM1 quantum singularity.
It must be noticed that a strong dependence of $T_{\rm CEP}$ quite similar to that observed here was found in a band structure frame with a Van Hove singularity.\cite{Binz}
Thus FS instability appears a key ingredient to discuss its FM quantum instability and also in AF heavy fermion systems such as CeRh$_2$Si$_2$, CeIn$_3$, or YbRh$_2$Si$_2$.

In summary, we have measured the ($P,T,H$) phase diagram of UGe$_2$ up to the vicinity of the QCEP using the Hall resistivity.
The QCEP of PM - FM boundary is suggested to be located at $3.5-3.6$ GPa and $17-19$ T, where it is sufficiently apart from the FM critical point at zero field to be separately observed ($P_{\rm QCEP}/P_{\rm C}>2$).
A strong evolution of the field dependence of the $A$ coefficient was observed from a strong first order regime up to the QCEP, while the Hall resistivity showed that the change in the FS at the PM--FM1 transition is maintained up to the vicinity of the QCEP.

\section*{Acknowledgement}

We acknowledge H. Tou, H. Harima, T. D. Matsuda, L. Malone, D. Braithwaite, J. P. Brison, H. Yamada, V. Mineev and M. Imada for supporting our research and valuable discussions.
This work was financially supported by the French ANR project (CORMAT, SINUS and DELICE) and ERC starting grant (NewHeavyFermion). J.F. is directeur de Recherche emerite in CNRS

\end{document}